\documentclass[useAMS,usenatbib,referee]{mn2e}
\usepackage{epsfig}
\usepackage{graphicx}

\begin{document}
\voffset=-0.5 in


\title[Dynamical formation of accreting IMBHs]
{On the dynamical formation of accreting  intermediate mass black holes}

\author[A.G. Kuranov, S.B. Popov, K.A. Postnov, M. Volonteri, R. Perna]
{A.G. Kuranov $^{1}$, S.B. Popov $^{1}$, K.A. Postnov $^{1,2}$,
M. Volonteri $^3$  and R. Perna$^{4}$
\thanks{E-mail: alex@xray.sai.msu.ru(AGK); polar@sai.msu.ru (SBP)}\\
$^1${\sl Sternberg Astronomical Institute, Universitetski pr. 13,
Moscow, 119992, Russia} \\
$^2${\sl Department of Physics, Moscow State University, 119992, Russia}\\
$^3${\sl Northwestern University, Department of Physics and Astronomy, 2131 Tech Drive, Evanston, IL, 60208, USA}\\
$^4${\sl JILA and Department of Astrophysical and Planetary Sciences, University of Colorado, 440 UCB, Boulder, CO, 80309, USA}\\ }
\date{Accepted ......  Received ......; in original form ......
      }

\maketitle

\begin{abstract}
We compute the probability that intermediate mass black holes (IMBHs)
capture companions due to dynamical interactions and become accreting
sources, and explore the possibility that the accreting IMBHs would
appear as ultra-luminous X-ray sources (ULXs).  We focus on IMBHs
originating from low-metallicity Population III stars. Two channels of
IMBH formation are considered: from primordial halos in the framework
of hierarchical clustering, and from non-mixed, zero-metallicity
primeval gas in galactic discs. IMBHs can form binary systems due to
tidal captures of single stars and exchange interactions with existing
binary systems in galactic discs.  We find that neither formation
mechanism of the accreting IMBH binary is able to provide enough
sources to explain the observed population of ULXs. Even at sub-ULX
luminosity, the total number of accreting IMBHs with $L> 10^{36}$
erg~s$^{-1}$ with dynamically captured companions is found to be $<
0.01$ per galaxy.
\end{abstract}


\section{Introduction}
Intermediate mass black holes (IMBHs) are a class of objects with masses
larger than standard stellar-mass black holes (BHs), but much smaller
than the $\sim10^6-10^9\,M_\odot$ of supermassive black holes detected
in the centers of galaxies, including our own Milky Way.
The masses of stellar
BHs are expected theoretically to span a range
between
roughly $\sim5-50\,M_\odot$ \citep{Fryer2001}.
These stellar-mass BHs are the end points of
stellar evolution for
sufficiently massive
stars formed out of metal enriched gas. At {\it
zero metallicity\/}, much more massive stars and black holes may be
forming. Indeed it has been suggested that
zero-metallicity stars can
form with masses of $\sim10^2-10^3\,M_\odot$, and produce IMBHs
directly
as a result of stellar evolution (\citealt{heger2003}).
IMBHs can be remnants of zero-metallicity (Population III) massive
stars of different origin.  The first possibility is that IMBHs are
produced by primordial Population III stars at early stages of galaxy
formation and some could be left wandering in galactic halos, as
suggested by numerical simulations (see \citealt{vhm2003};
\citealt{metal2004}; \citealt{vp2005} and references therein). Due to
the so-called gravitational rocket mechanism \citep{metal2004,
MadauQuataert2004, Merrittetal2004, Blanchetetal2005, Bakeretal2006},
IMBHs could receive kick velocities high enough to prevent their
merging into the central galactic BH, but small enough not to leave
the potential well of the Galaxy.  Calculations by \cite{vp2005}
indicate that the number of such ``relic'' IMBHs in the internal parts
of a typical galactic halo can amount to a few dozens per galaxy. The
second scenario is based on the recent suggestion by \cite{jh2006}
that about 10-30\% of stars in galactic discs at $z\sim3$~-~4 can be
born from primordial gas due to inefficient mixing of the interstellar
medium.  If evolution of the most massive of these stars leads to the
formation of IMBHs, they should be left inside the galactic disc and
have the low velocity dispersion typical of galactic disc
populations.

IMBHs can capture stellar companions due to dynamical interactions and
appear as accreting sources \citep{mr2001}.
For IMBHs formed in young, dense stellar clusters, these processes
were considered by \cite{Blecha&al06}. These authors concluded
that the  dynamical formation of binaries with accreting IMBHs, observable as
bright ultra-luminous X-ray sources (ULXs) in stellar clusters, is
very inefficient.
In this paper we further study this suggestion, focusing on the possible formation of ULXs
in binary systems as a result of dynamical
interactions of IMBHs with the galactic disc population.

ULXs are bright point-like objects observed
in external galaxies. They are considered to form
a separate class of objects apparently associated with young stellar
populations. ULXs are characterized by high X-ray luminosity
($L_{\mathrm{x}} \sim 10^{39}$~--~$10^{42}$ erg~s$^{-1}$) and often strong variability \citep{cm1999}.
The observed number of ULXs
is about 0.1 per galaxy for $L_{\mathrm{x}} \ga 2 \times
10^{39}$ erg~s$^{-1}$ and 0.01 per galaxy for $L_{\mathrm{x}} \ga 10^{40}$
erg~s$^{-1}$ \citep{cm2005}.

The nature of ULXs is unclear.
Two competitive hypothesis (see \citealt{m2004} for a recent review)
suggest that either these sources are accreting
binary systems with stellar-mass black holes and significant
collimation of radiation (like SS433 in our Galaxy) seen along the jet axis
\citep{fm2001, ketal2001, bkp2006},
or they are manifestations of
intermediate mass black holes with $M\sim 100-1000 M_\odot$
\citep{cm1999, mr2001}.
The first hypothesis is supported by observations of emission nebulae around
some of these sources similar to that observed around SS433
(see \citealt{f2006}
and references therein for more details),
while the IMBH hypothesis is advocated by observations of $\sim 20$ s
quasi-periodic X-ray oscillations in at least one source \citep{sm2003}.
Observations of specific X-ray spectral features forming in the funnel
of supercritical accretion disc seen along the jet
may be crucial in distinguishing
between these two possibilities \citep{f2006}.
For ULX sources which do not display a rapidly variable luminosity, another possible
model is that of very young, fast rotating pulsars,
whose X-ray luminosity is powered by their rotational energy (Perna \& Stella 2004).

If accretion is the source of energy, the observed X-ray luminosity
$>10^{39}$ erg~s$^{-1}$ suggests an accretion rate onto the compact
star of $>10^{-7} M_\odot$~yr$^{-1}$ (assuming 10\% efficiency).
Volonteri \& Perna (2005) investigated the possibility that ULXs can
be associated to wandering IMBHs which accrete material from the
surrounding interstellar medium (see also Islam, Taylor \& Silk 2004;
Mii \& Totani 2005; Mapelli, Ferrara \& Rea 2006). These calculations
showed that the typical accretion rate is too low to produce bright
ULXs, unless the wandering IMBHs carry a substantial baryonic core
with them (e.g. the stripped remnants of the original host galaxy of
the wandering BH).  

A different route which could provide a high accretion rate for IMBHs
is mass transfer from a companion at an advanced evolutionary stage.
There are two possibilities for an IMBH to capture a companion star:
tidal interactions with single stars and exchange interactions with
binary systems. Tidal interactions are efficient for sufficiently
close fly-bys (at a distance of the order of a few stellar radii), so
the tidal capture rate is fairly small even for low velocity
dispersion in the galactic disc; in addition, the tidal destruction of
the entire captured star is very likely.  The advantage of tidal
captures is that sufficiently close binaries with IMBHs are formed and
are likely to become bright X-ray sources.

In the case of exchange interactions, the characteristic cross-section
is larger than for tidal captures -- it scales in proportion to the
binary semi-major axis.  This increases the probability of interaction
between an IMBH and the binary star.  On the other hand, {\it i)} for
large relative velocities between IMBHs and binaries, only a small
fraction of sufficiently close systems (called ``hard'', see below)
can produce bound binaries after interaction, thus decreasing the
effective cross-section;  {\it ii)} after the exchange of the
components, the binary's semi-major axis increases thus decreasing the
chances for the IMBH in such a binary to become a bright X-ray source.

In this paper we study both processes of formation and evolution of
binaries with IMBHs in galactic discs
using population synthesis calculations. High-velocity halo IMBHs
will be considered, as well as low-velocity IMBHs formed in the galactic
disc.

\section{Model}

Our model includes the following steps.
\begin{itemize}
\item We specify the population of IMBHs and its properties.
For halo IMBHs we use the spatial distribution and velocities
calculated by \cite{vp2005} as initial conditions, and
keep pace with their trajectories in the galactic
potential over  $10^{10}$~yrs,
selecting the crossings of the disc plane.
For disc IMBHs, we follow the proposal by \cite{jh2006} and consider
their formation from massive stars as described below (\S2.2).
\item
We calculate the exchange interactions of IMBHs with different binary systems or
tidal captures of single stars in the disc.
\item Using a modified version of the SCENARIO MACHINE -- the binary population
synthesis code \citep{lpp1996} --
we calculate the subsequent evolution of captured systems.
\item Summing up all probabilities for a given IMBH to appear as ULX
in a binary formed during all disc crossings, and making appropriate normalizations,
we calculate the expected number of ULXs in a typical spiral galaxy
 with assumed constant star formation.
\end{itemize}

\subsection{IMBHs wandering in halos}
The initial spatial and velocity distributions of IMBHs are taken from the numerical
calculations by \cite{vp2005}.
Coordinates and velocities of these IMBHs have been taken from a stationary
distribution corresponding to the present time (i.e. to the redshift
$z=0$).
We have selected only those IMBHs which were found inside the region bounded by
galactocentric distance  $R<15.4$~kpc in the galactic plane
and height $|Z|<3$~kpc over the galaxy plane.  
This choice reflects the model density
profile adopted for the Galaxy stellar disc (exponential disc,
with scale length 7.7 kpc and scale height 1.53 kpc). With this parameter  choice, 
the stellar density  at $|Z|>  3$ kpc becomes extremely small, so that the probability 
of capturing a companion is negligible. Our test demonstrated that IMBHs
initially outside  $|Z|<3$~kpc contribute very little to the number of
galactic plane crossings.

The number of IMBHs within this spatial boundaries in a Milky Way sized galaxy strongly depends on the
assumed density peak cut during hierarchical clustering.
If IMBH formation happens only in rare  3.5~$\sigma$  density peaks, the
average number of wandering IMBHs with  $R<15.4$~kpc and  $|Z|<3$~kpc is 1.3
\citep{vp2005}. We also made a run for IMBHs originated
from more common 3$\sigma$ peaks. In that case there are on average 126 IMBHs per
galaxy.

After the initial velocity and position of an IMBH are selected,
we let the IMBH orbit to evolve in the galactic potential.
The motion of each IMBH is traced for
$10^{10}$~yrs with a time step of $10^4$~yrs.
The large integration time is chosen to increase the number of
galactic disc crossings, as we have only several dozens
IMBHs from direct numerical simulations; with the
typical IMBH orbital period of several $10^6$ yrs, we expect up to several
thousand crossings in our calculations. This artificial increase in the
number of trials helps to
obtain statistically significant results when calculating
dynamical interactions with stars. The final results are normalized appropriately.

Trajectories of IMBHs are calculated in a way similar to that used
in calculations of the spatial evolution of isolated neutron
stars in our Galaxy (see, for example,  \citealt{p2005} and references therein).
We made use of the axisymmetric galactic potential initially  proposed by
\cite{mn1975}. The potential includes three parts: disc, spheroid and
halo, which are described in detail in \cite{bm1993}.
IMBHs mostly cross the galactic plane at small ($<0.5$~kpc)
galactocentric radii (see Fig. 1). Spatial velocities of IMBHs
at each galactic plane crossing 
are collected and used for the calculation of
exchange interactions with binaries and tidal captures of stars.
Typical orbital periods of these IMBHs are about $10^7$ years, while
their velocities at the galactic plane crossings are $\sim$ 300-400
km~s$^{-1}$.
\begin{figure}
\includegraphics[width=0.45\textwidth]{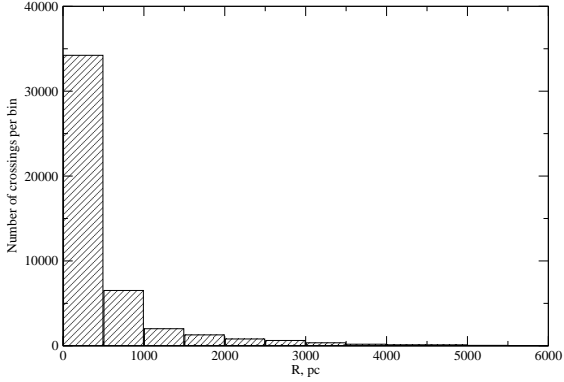}
  \caption{Distribution of the distances from the galactic center at which
halo IMBHs (from the 3.5$\sigma$ peaks)  cross the galactic plane.
}
\label{rdistr}
\end{figure}

\subsection{IMBHs from low-metallicity gas}

According to \cite{jh2006}, a fraction of about 10-30\% of gas in galaxies can preserve
primordial (low-metallicity) composition at $z\sim 3$~--~4.
The most massive stars formed from such a pristine gas can
produce IMBHs at the end of evolution.
Adopting a galactic star formation rate of 1 $M_\odot$ per year for $10^8$ yrs (as in \citep{jh2006}), and
using 0.1 for the fraction of primordial gas content, we obtain that
$10^7\, M_{\odot}$
of gas in the galaxy is transformed into zero-metallicity stars
at $z=3-4$.

To estimate the number of IMBHs formed in this way, we need to specify the initial mass
function, the mass interval(s) in which BHs are formed, and the BH masses.
The initial mass function is taken in a power-law form:
\begin{equation}
\label{m_bh}
f(M)~\propto M^{-\alpha}.
\end{equation}
Two values of the index $\alpha$ were considered:
2.35 (the classical Salpeter value) and
$\alpha=1.35$. 
The second choice is motivated by the possibility of a significant
flattening of the initial mass function for very massive stars 
 which can
lead to the increase of the number of ULXs in simulations.
 For example, the extreme
variant $\alpha=0$ (flat initial mass function)
was discussed, for example, by \cite{cdc1995}.
The mass range of zero-metallicity stars which evolve into black holes
was chosen to consist of two intervals:
from 30 to 140 $M_{\odot}$ and from 260 to  1000 $M_{\odot}$~\citep{whw2002}.
Inside the first mass interval, we assumed $M_{\mathrm{BH}}=0.5M_{\mathrm{star}}$, while
inside the second interval $M_{\mathrm{BH}}=0.75M_{\mathrm{star}}$. This specific choice,
however, does not affect our results significantly.  Neither does the choice of the upper
mass limit (for example, 500~$M_{\odot}$ as in \citep{op2003})
because the stellar initial mass
function rapidly decreases towards higher values.

The velocity distribution of disc IMBHs was assumed to be 
Maxwellian with a characteristic velocity dispersion of 10
km~s$^{-1}$, which is typical for young massive stars in the Galactic
disc. We assume that the velocity dispersion remains constant with time and do not
trace the individual motion of such IMBHs. This means that we neglect
the possible dynamical heating of the IMBH population; any increase of 
the velocity dispersion, however, would reduce the efficiency of the dynamical 
interactions of IMBHs with stars.

\subsection{Formation of binaries due to exchange interactions}

The probability of a dynamical interaction  $p$
between an IMBH and a (binary) star in a time interval $\Delta t$ is:
\begin{equation}
p= 1-\mathrm{exp}\left\{ -\frac{\mathrm{\Delta t}}{\tau}\right\},
\end{equation}
where $\tau=(\mathrm{n}\mathrm{\langle V \rangle}\sigma)^{-1}$
is the characteristic time of interaction,
$n$ is the local stellar density, $\sigma$ is the cross section of the interaction, and
$V$ is the relative velocity of IMBHs and the (binary) star at infinity.

We have used two fitting formulas for the exchange cross section:
one proposed by \cite{h1991} and one by \cite{hhm1996}.
\cite{h1991} simulated encounters between ``hard''
binaries with equal mass binary components ($M_0$)
and intruders with extreme masses relative to the masses of the components.
In the limiting case of large IMBH mass ($M_{\mathrm{IMBH}}\gg M_0$),  the mean time
($\tau$) of the exchange interaction can be estimated as
\begin{equation}
\label{hills}
\tau_{}=1.9\times 10^{14}{\mathrm {yr}}\left(\frac{1/{\mathrm {pc}}^3}{n}\right)
\left(\frac{\langle V^2 \rangle^{1/2}}{10~{\mathrm{km/s}}}\right)
\times
\end{equation}
$$
\left(\frac{{\mathrm{A.U.}}}{a_0}\right)\left(\frac{0.5M_\odot}{M_0}\right)\left(\frac{M_0}{M_{\mathrm{IMBH}}}\right)^{5/6}\;,
$$
where $a_0$ is the semimajor axis of the binary.

\cite{hhm1996} calculated the exchange cross section in the limit of
very low-speed intruders only.
These authors, however, expanded their calculations for a wide range of mass ratios.
The equation obtained by \cite{hhm1996}
for the exchange cross section  has a complicated dependence
on the masses of all interacting systems.
  Let $x =
m_1/M_{12}$,  $y = m_3/M_{123}$ and $M_{ij} = m_i + m_j$,
where $m_1$ is the ejected component, $m_3$ is IMBH.  Then
\begin{equation}
\label{hut}
\sigma  \simeq  {GM_{123}a\over V^2}{m_3^{7/2}M_{23}^{1/6}\over
M_{12}^{1/3} M_{13}^{5/2}M_{123}^{5/6}}
\exp (3.70 + 7.49x -1.89 y 
- 15.49x^2 -
\end{equation}
$$ 
- 2.93xy - 2.92y^2 + 
3.07x^3 + 13.15x^2y
-5.23xy^2 + 3.12y^3).\\
$$

This formula fits 75\% of the numerical calculations to better than 20\%; larger
discrepancies are restricted to mass ratios where the cross sections are
probably too small to be of importance in applications.  Hut et al. have also
compared their formula with results obtained by other authors, and found that the
agreement is generally satisfactory. 
The results of 
our calculations presented below also do not depend significantly on the 
details in the formula for the exchange cross section,  i.e., we obtain
similar numbers of ULXs using  Eq.~(3) and  Eq.~(4).


\subsection{Tidal capture}

IMBHs should tidally interact with single stars, too.
A detailed description of the dissipation in the
tidal capture process can be found in \cite{fetal1975}.
The tidal capture cross section can be computed from the requirement that the
deposition of
kinetic energy during the encounter, $\Delta E_T$, is larger than
the relative energy of the pair at infinity: $E=\mu v_{\infty}^2/2$.
Here $\mu=M_1M_2/(M_1+M_2)$ is the reduced mass of the pair of stars with masses
$M_1$ and $M_2$.

It is interesting to compare the tidal cross section to the exchange
interaction one. As the gravitational focusing dominates in both cases, the
cross section ratio is given by the formula (Hills 1992):
\begin{equation}
\frac{\sigma_{\mathrm{tidal}}}{\sigma_{\mathrm{exch}}}\simeq 2.5
\frac{R_\mathrm{s}}{a_0},
\end{equation}
where $R_{\mathrm{s}}$ and $a_0$ are, respectively, the radius of the star and the semimajor axis of the
pre-encounter binary.
In our calculations, however, we use a more
precise formula for the tidal capture cross section from \cite{kl1999}, rather
than the simple scaling given above.
These authors present a convenient fitting formula for the cross section of
tidal capture for two stars with a large mass ratio as a
function of their relative velocity at infinity.

\cite{mr2001}
estimated the tidal capture rate of massive IMBHs (Population III
remnants) at the Galactic center. They noted that, although the capture rate in
dense stellar regions is high enough, because of the high velocity dispersion
 the close encounters leading to the tidal disruption of
 the main-sequence star will be very common, leading to a consistent decrease in the
 estimate of surviving binary systems which can appear as ULXs.

\subsection{Binary population synthesis}

We used the population synthesis method
to take into account different types of binaries
interacting with IMBHs.
Specifically, the modified version of the SCENARIO MACHINE
code  by \cite{lpp1996} was adopted \footnote{See the
on-line material at
http://xray.sai.msu.ru/$\sim$mystery/articles/review/.}.

{\em i) Halo IMBHs.} In this case, for each IMBH and for each
galactic plane crossing, we randomly selected $N$ binary
systems whose evolution is calculated according to
standard initial distributions (see below).
Since each IMBH crosses the galactic plane
$ N_\mathrm{cross}\sim 10^4$ times
during $10^{10}$~yrs, the total number of potential encounters
with binaries
is  $N_{\mathrm s}= N\times  N_{\mathrm {cross}}$.
Statistically significant results are obtained for $N_{\mathrm s}\sim 10^5$,
so in our calculations we selected several tens of
binaries per each IMBH. A few thousand bright X-ray sources
typically appeared in a population synthesis run
consisting of $10^5$ binaries with IMBHs.

{\em ii) Zero-metallicy stars.} Similarly, to obtain statistically significant results,
for each IMBH originated from zero-metallicity stars in the disc we modeled
$N \sim 10^5$ binary systems.

\subsubsection{Initial binary distributions}

Initial distributions of binary parameters are taken according
to standard premises widely used
in the population synthesis calculations (\citealt{Popova&al82}).

{\em a)} Orbital separation $a$.

We use equprobable distribution in the logarithmic scale.            
\begin{equation}
\label{loga}
dN/d(\log a)={\mathrm const}, ~~~~~~ \max \left\{
\begin{array}{lcl}20R_\odot\\
\rm{R_L(M_1)}\\
\end{array}
\right\}
<a<10^4  R_\odot,
\end{equation}
where $R_{\mathrm L(M_1)}$ is the radius of the Roche lobe of the primary component.
The lower limit is determined by the condition of coalescence of the
components at the main sequence. The choice of the upper limit is dictated 
by the following considerations. First, wide binaries (with semimajor axes 
up to $10^7  R_\odot$) are soft and will be disrupted by dynamical 
interactions with an IMBH, so it is not necessary to compute their evolution 
in detail. Hence we used the upper limit $10^4 R_\odot$ 
for initial binary semi-major axis. This value is also close to the boundary
beyond which binaries will be destroyed by dynamical interactions in stellar
systems with velocity dispersion of around 100 km/s, which is typical for 
central regions of galaxies.  
The account in the statistics  for wide binaries with $10^4<a/R_\odot<10^7$, 
which can be present in galactic disks, 
is easily made as we use flat probability distribution in the logarithmic
scale. For correction we just need to divide 
the calculated rates by a numerical coefficient of the order of two as we
took binaries from one half (from $\sim \log a=1$ up to $\log a=4$) of the full
range (from $\log a=1$ up to $\log a=7$).

{\em b)} Masses of the components.
We use the Salpeter  mass function
for the mass of the primary (i.e. more massive) star:
\begin{equation}
\label{m1}
dN/dM_1 \propto M^{-2.35}_{1},~0.1M_\odot<M_1<120M_\odot.\\
\end{equation}

{\em c)} Mass ratio of the binary components~$q=M_2/M_1\le1$.
\begin{equation}
\label{m2}
dN/dq~\propto q^{\alpha_{\mathrm q}}\,.
\end{equation}
We present results for $\alpha_{\mathrm q}=1$, i.e. components with equal masses
are more probable.
This value is favored by some population synthesis analysis
(e.g. \citealt{lpp1996}).  Calculations were also made for $\alpha_{\mathrm
q}=0$ (flat mass ratio distribution).
For this value we obtain a smaller number of bright X-rays sources with IMBHs,
but the difference is less than a factor of two.

{\em d)} Initial orbital eccentricities are assumed to be zero. This
simplification is justified by the fact that, for the close
binaries of interest, tidal circularization occurs on time
scales much shorter than other characteristic evolutionary times.
The number density of stars in the galactic disc is assumed to be
one per cubic parsec. This value is an order of magnitude higher
than that in the solar vicinity, but it should be typical for
central regions of galaxies, in which we are interested
\citep{ms1996}. Our results on the bright source statistics can,
however, be easily generalized to other values of the stellar
density since they scale linearly with it.

The disc (for the case of wandering IMBHs) is modeled as a flat
structure with semithickness of 300 pc.  Since most crossings happen
at small galactocentric radius, such a thick disc is a reasonable
approximation.  We assume that all stars belong to binaries.  As the
fraction of massive stars in binary systems is not less than 50\%,
this appears to be a safe assumption.

\subsubsection{Formation of binaries with IMBH}

The evolution of a binary system is considered as a sequence of
stages, and the duration of each one is defined by the shortest evolutionary
timescale of one of the binary components.
The states of stars and binary parameters at the beginning of a stage
uniquely determine the fate of the binary up to the end of the stage.

At each evolutionary stage $i$
of the binary  $j$
($j=1..N, N\sim 10$ for halo IMBHs
and $N\sim10^5$ for IMBHs originated from low-metallicity stars),
we calculate the probability of the  exchange
interaction $p^{ij}_{\mathrm{capt}}$
with an IMBH during the BH galactic plane crossing time
$t_{\mathrm{cross}}$ for the case of wandering IMBHs:
\begin{equation}
p^{ij}_{\mathrm{capt}}= \, \frac{t_{\mathrm{cross}}}{\tau^{ij}}\,.
\end{equation}
For IMBHs inside the disc originated from zero-metallicity stars this
probability reads:
\begin{equation}
p^{ij}_{\mathrm{capt}}= \, \frac{t_{\mathrm{Gal}}}{\tau^{ij}}
\end{equation}
Here $t_{\mathrm{Gal}}=10^{10}$~yrs is the galactic age.

\subsubsection{Evolution of binaries with IMBH}

After an IMBH acquires a   companion, we calculate the
new semimajor axis $a_\mathrm{f}$ of the newborn binary system.
For exchange interactions, $a_\mathrm{f}$ is calculated
according to \cite{h1991}. For $M_\mathrm{IMBH}\gg M_0$ the
harmonic mean semimajor axis of the post-encounter binary for all exchange
collisions in the limit of hard binaries is:
\begin{equation}
a_\mathrm{f}=12\, a_0\left( M_{\mathrm{IMBH}}/100 M_{\odot} \right) ^{2/3},
\label{af}
\end{equation}
where $a_0$ is the semimajor axis of the binary system before
the exchange interaction.
As it was shown in  \cite{h1991}, the distribution of $a_\mathrm{f}$ has the
following property.
For $ M_{\mathrm{IMBH}}/M_0 =1000$ (for which $a_\mathrm{f}/a_0=56$),
there are almost no binaries in the semimajor axis range
$a_\mathrm{f}\sim (1-20)a_0$.
The number of binaries rises rapidly for semimajor axis
$a_\mathrm{f}>50a_0$. That is why the use of Eq.~(\ref{af}) does not
underestimate significantly the number of accreting systems formed due to
exchange interactions.

For binaries formed through tidal captures
we assume that all captured stars survive in a circular orbit around the IMBH at
$r=2 r_\mathrm{t}$, where $r_\mathrm{t}=(M_\mathrm{BH}/m)^{1/3} R$
is the tidal-breakup radius.

After the binary parameters are determined, we calculate the
evolutionary track of the system using the population synthesis code.
Calculations continue up to the age $t=10^{10}$~yrs.  The statistics
of accreting binary sources is collected for all halo IMBHs and disc
IMBHs.

\subsection{Bright sources statistics}

Since we calculate the probability to capture a companion from a
given (randomly selected) binary system, it is necessary to take into
account two effects. The first
regards the duration, $t^{ij}_{\mathrm{stage}}$, of the stage $i$
at which the system $j$ was taken. The second is due to the fact that the ULX
stage ($t^{ij}_{\mathrm{ULX}}$) is just a short episode in the life of
the formed system.
 We assume that an accretion rate $\dot{M}$ yields a bolometric
luminosity $L_\mathrm{bol}=\eta\dot{M}c^2$, where
$\eta\sim 0.1$ is a typical value of the
accretion efficiency \citep{frank92}.
In Figs.~2-3 and 5-6 we
have marked the value $\dot{M}\sim 10^{-6} M_{\odot}$~yr$^{-1}$, which
corresponds to a bolometric luminosity of $L_\mathrm{bol}\sim
10^{40}$ erg~s$^{-1}$. This is the value we will be referring to
when computing the statistics of ULXs.
Since the processes we study here yield very small
probabilities for the observation of bright sources, we limit
our results to the bolometric luminosities, keeping in mind that
these in turn provide an upper limit to the X-ray luminosities.

Given our definition of ULX, for each crossing $k$ of the galactic
plane by an IMBH, we obtain the probability that it can shine as a
ULX:
\begin{equation}
p_{\mathrm{k}}= \frac{1}{\mathrm{N}} \sum_{j=1}^\mathrm{N} \sum_{i} p^{ij}_{\mathrm{capt}} \, \frac{t^{ij}_{\mathrm{stage}}}{t_{\mathrm{Gal}}}
\,  \frac{t^{ij}_{\mathrm{ULX}}}{t_{\mathrm{Gal}}}.
\label{threeprobs}
\end{equation}
Here $t_{\mathrm{Gal}}$ is the galactic age, and
$t^{ij}_{\mathrm{stage}}$ is the duration of the stage at which the donor was
captured.

The probabilities $p_{\mathrm k}$ given above refer to each IMBH used in our calculations.
We sum these probabilities in each galactic plane crossing by all IMBHs,
and then divide the result by the number of merging trees
from which these IMBHs were selected.
The final statistics of the number of observable bright X-ray
sources can be obtained as follows:
\begin{equation}
\mathrm{N}_\mathrm{systems}=\frac{1}{\mathrm{N}_\mathrm{gal}}
\sum_1^{\mathrm{N}_{\mathrm{BH}}} \sum_k^{k_{\mathrm{max}}}
p_{\mathrm k}.
\label{ntotal}
\end{equation}
Here $N_\mathrm{gal}$ is the number of galaxies (merging trees). For
$3.5\sigma$ halos $N_\mathrm{gal}=30$, for $3\sigma$ halos
$N_\mathrm{gal}=1$. The number of IMBHs, $N_{\mathrm{BH}}$, is equal to 40 in the case
of $3.5\sigma$ halos, and 126 for  $3\sigma$ halos. The total number of
galactic plane crossings for each IMBH, $k_{\mathrm{max}}$, is calculated explicitly and
is of order of $10^3$.

\section{Results for exchange interactions}

In this section we present the results of our calculation for exchange interactions.

\subsection{Halo IMBHs}

In Fig.~2  we show the number of  accreting binaries with IMBHs
per galaxy for 3.5$\sigma$ halos (dotted and dashed lines), and in
Fig.~3 for 3$\sigma$ halos. It can be seen that the probability to
form an ULX through exchange interaction of a halo IMBH with
binaries in the galactic disc is extremely small.

In particular, for the cross section from \cite{hhm1996}, which is
more realistic in our case, we obtain that the probability to observe
a source with $\dot M > 10^{-6} \, M_{\odot} \, {\mathrm{yr}}^{-1}$ is
only $\sim 3 \times 10^{-14}$ per per galaxy.  For the Hills' cross
section the result is even smaller: $\sim 1.8 \times 10^{-14}$.

Relaxing the threshold on the accretion rate for an ULX increases the
number of sources, but insignificantly: e.g. for an accretion rate of
only $10^{-10} \, M_{\odot}$~yr$^{-1}$, we get $N_{\mathrm{ULX}}\simeq 10^{-9}$
per galaxy.  Therefore, the number of accreting X-ray sources of any
luminosity due to ``primordial'' IMBHs capturing companions via
exchange interactions with binaries in the galactic disc is
negligible.

\begin{figure}
\includegraphics[width=0.7\textwidth]{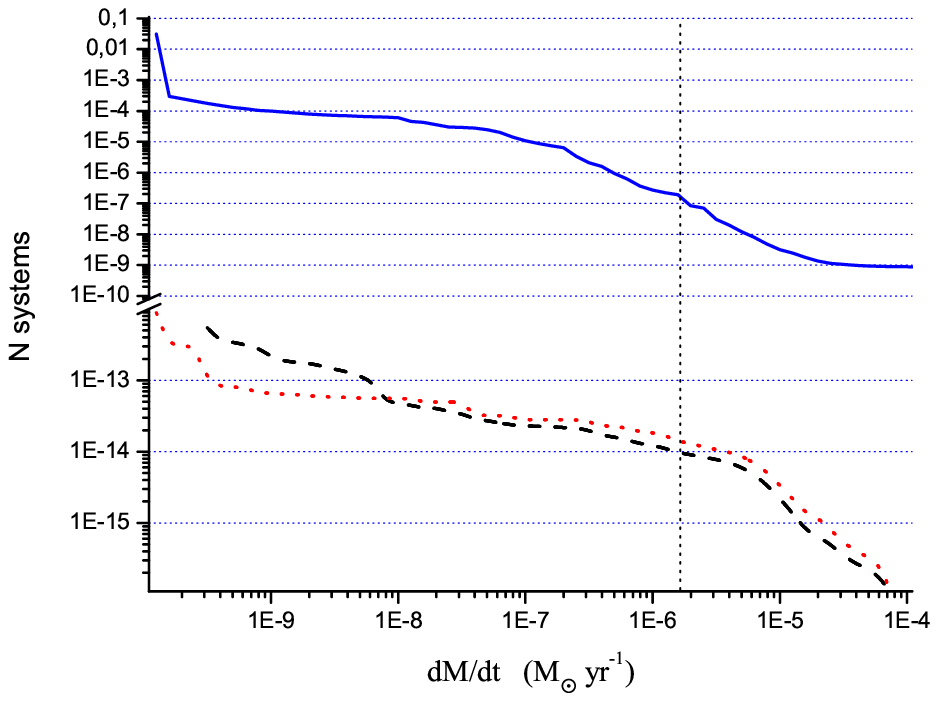}
\includegraphics[width=0.7\textwidth]{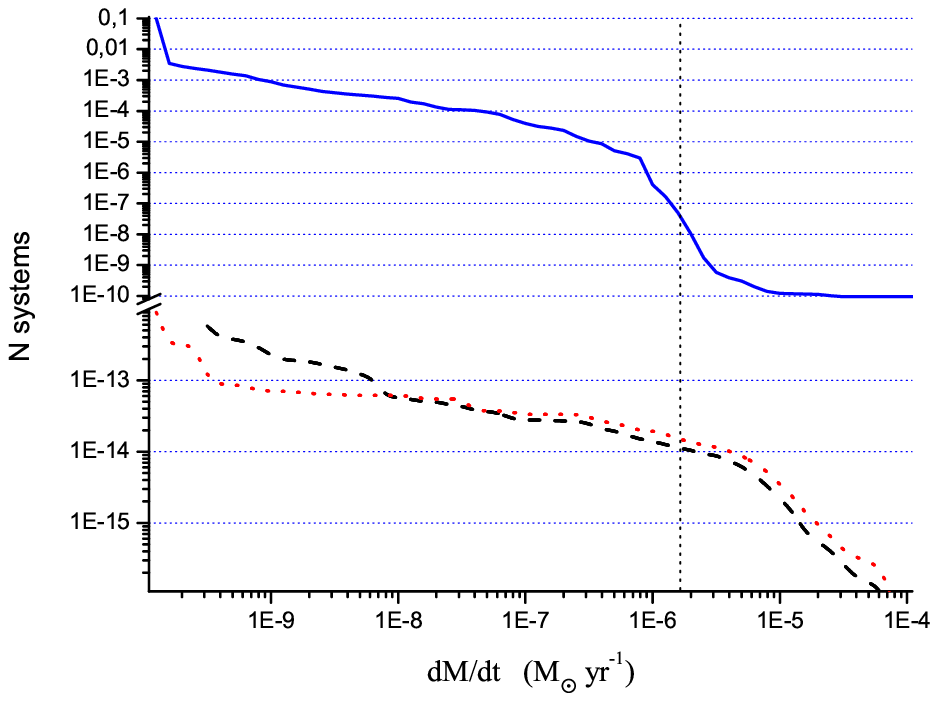}
  \caption{Integral distribution of the accretion rate for BHs
wandering in a galactic halo.  IMBHs born in 3.5$\sigma$ halos are
taken into account.  {\em Solid line}: systems formed via tidal captures.
{\em Dotted line}: systems formed via exchange interaction (using
Hut, Heggie \& McMillan 1996 fitting formula for the cross section).
{\em Dashed line}: systems formed via exchange interaction (Hills
1991 fitting formula).  The vertical dotted line shows the accretion
rate, corresponding to the luminosity $L_\mathrm{bol}=
10^{40}$~erg~s$^{-1}$, where $L_\mathrm{bol}=0.1 \dot M c^2$. {\em Upper panel}:
the duration of accretion is equal to the thermal time ($\sim 10^4$ yrs) of the
red giant's envelope.  {\em Bottom panel}: the duration of accretion is equal to
the He-burning time in the stellar core.
}
\label{halo35}
\end{figure}

\begin{figure}
\includegraphics[width=0.7\textwidth]{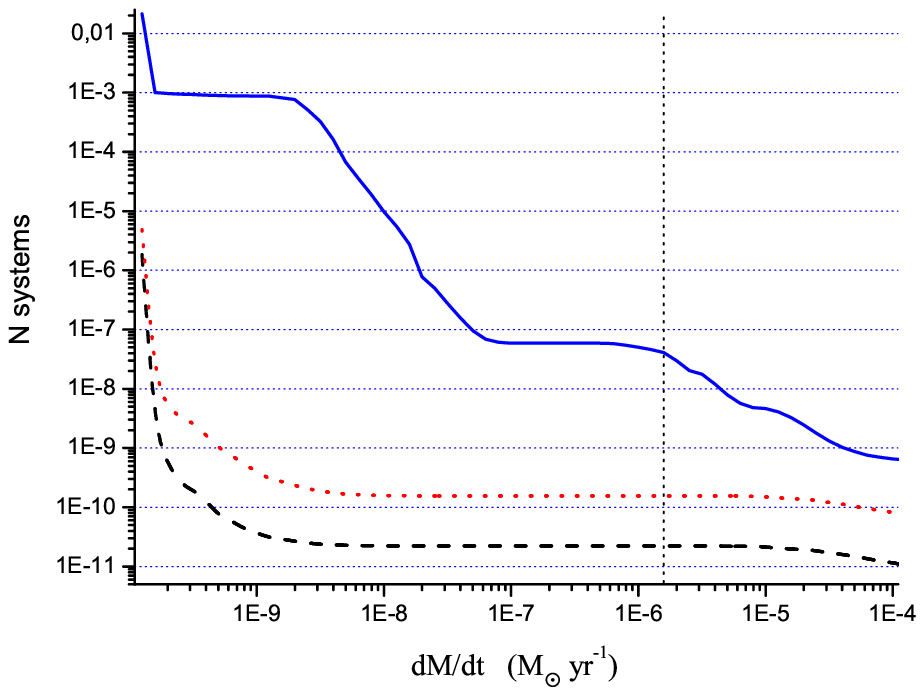}
\includegraphics[width=0.7\textwidth]{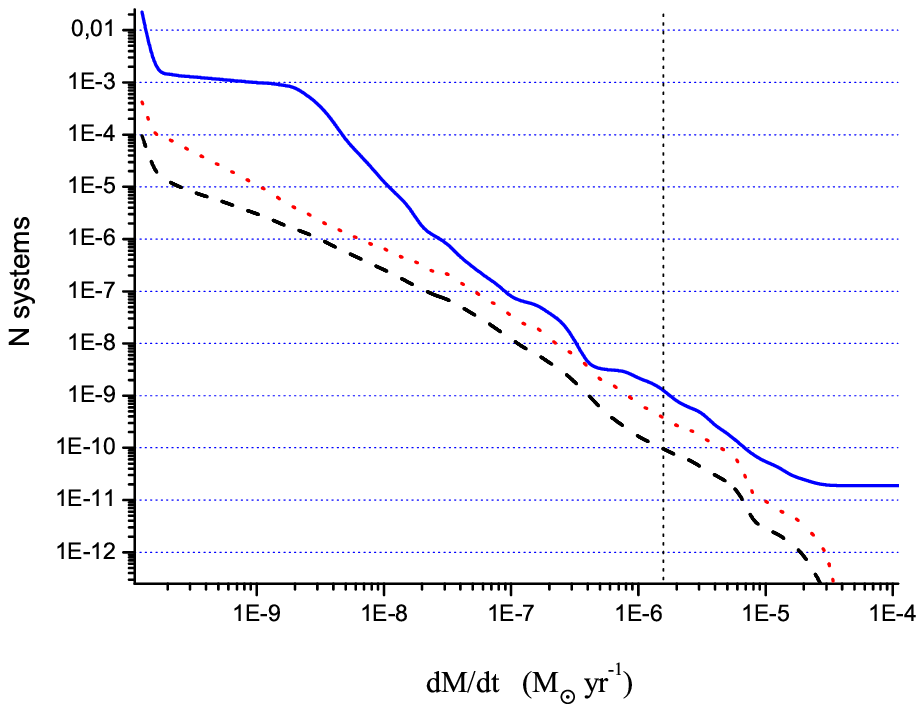}
  \caption{Integral distribution of the accretion rate for IMBHs wandering in the galactic halo.
IMBHs from 3$\sigma$ halos are used. Lines and panels as in Figure \ref{halo35}.}
\label{halo30}
\end{figure}

In our calculations we found two main types of donors feeding IMBHs.
Most frequently the companions are red giants. In this case the accretion rate
is  $ > 10^{-6}\, M_{\odot}$~yr~$^{-1}$. The duration of this
stage should be rather small: $\sim 10^4$ yrs, as this time is determined
by the thermal time of the convective envelope.
Since the mass ratio in the IMBH binaries is very high,
we also performed calculations
under the assumption that the duration of the accretion stage
is determined by the helium burning time in the stellar core, which is
0.1 of the hydrogen burning time. Again, the results are not
significantly changed. Both results are shown in the Fig.~\ref{halo35}
for the case of 3.5$\sigma$ halos, and in Fig.~\ref{halo30}
for 3$\sigma$ halos.

In some instances we found accreting white dwarfs as
companions. In this case the accretion rate is smaller
($\dot M < 10^{-8}\, M_{\odot}$~yr~$^{-1}$), but the duration of the stage
can be much longer. Such systems are formed
if the semimajor axis of the binary {\it
before} the interaction with an IMBH is small.

\begin{figure}
\includegraphics[width=0.45\textwidth]{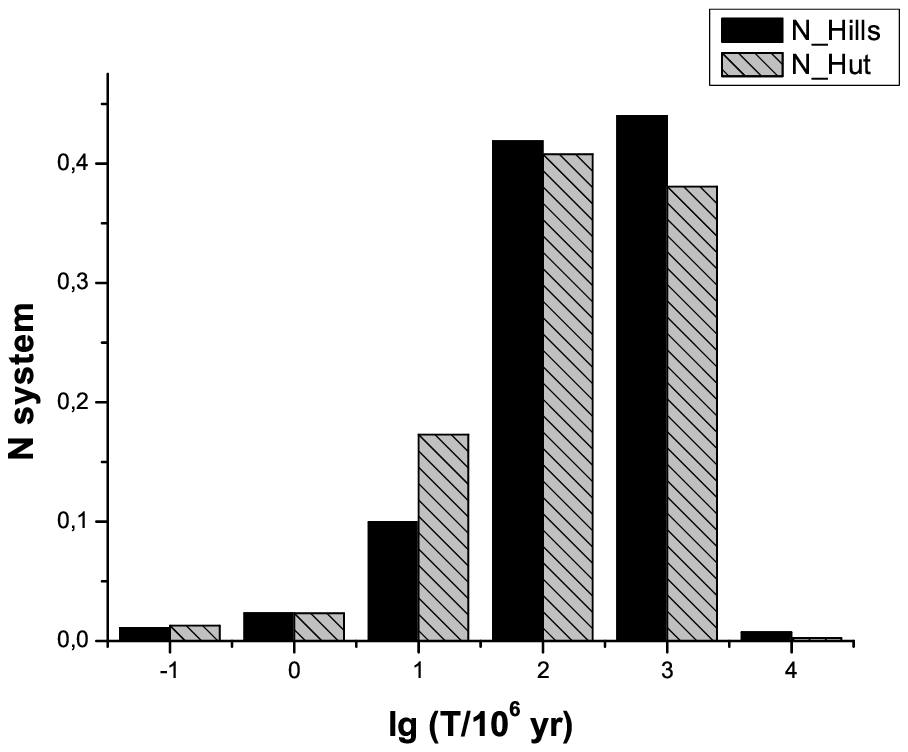}
  \caption{Total number of accreting IMBHs (i.e. independent of their
luminosity) vs. time of the on-set of accretion (time is counted since
the moment of formation of the binary system).  Calculations are
performed for IMBHs from 3.5$\sigma$ halos and assuming that the time
scale for accretion is set by the thermal time of the convective
envelope.  Time is in Myrs. Results are presented for two variants of
the cross-section according to Hills (1991) and Hut et al. (1991). For
each variant the sum in all bins is normalized to unity.}
\label{time}
\end{figure}

In Figure~\ref{time} we show the distribution of sources with $\dot M
> 10^{-6}\, M_{\odot}$~yr$^{-1}$ versus the time of the on-set of the
ULX stage.  For most systems this time is larger than the orbital
periods of the IMBHs, which are about $10^7$~yrs, since we chose for our
calculations only those IMBHs that spend most of their life close to
the galactic plane.  Because of that, most accreting systems follow the
IMBH spatial distribution.  Still, as the fraction of systems in
which accretion starts on timescale shorter than the orbital period
is non-zero, a slight concentration (on the level of few per cents)
towards the galactic plane (and center) appears.

\subsection{Disc IMBHs}

In the case of IMBHs originated in the galactic disc from
zero-metallicity massive stars we obtain a somewhat higher number
of ULXs $\sim 10^{-8}$ per galaxy (see Fig.\ref{disk235}).
A more effective
formation of accreting binaries in this case results from the
smaller velocity dispersion of the disc IMBHs as compared to the halo
ones. Still, this is not enough to explain observations. The
results of calculations are presented in Fig. \ref{disk235} for
the Salpeter initial mass function  of zero-metallicity stars
($\alpha=2.35$).
 In the case of $\alpha=1.35$ results are generally similar. 
The number of sources is several times higher, but still not enough to
explain the observed population of ULXs. Taking even flatter initial mass
function we can increase this number further more, but still not enough.

\section{Results for tidal capture}

\subsection{Halo IMBHs}

\begin{figure}
\includegraphics[width=0.7\textwidth]{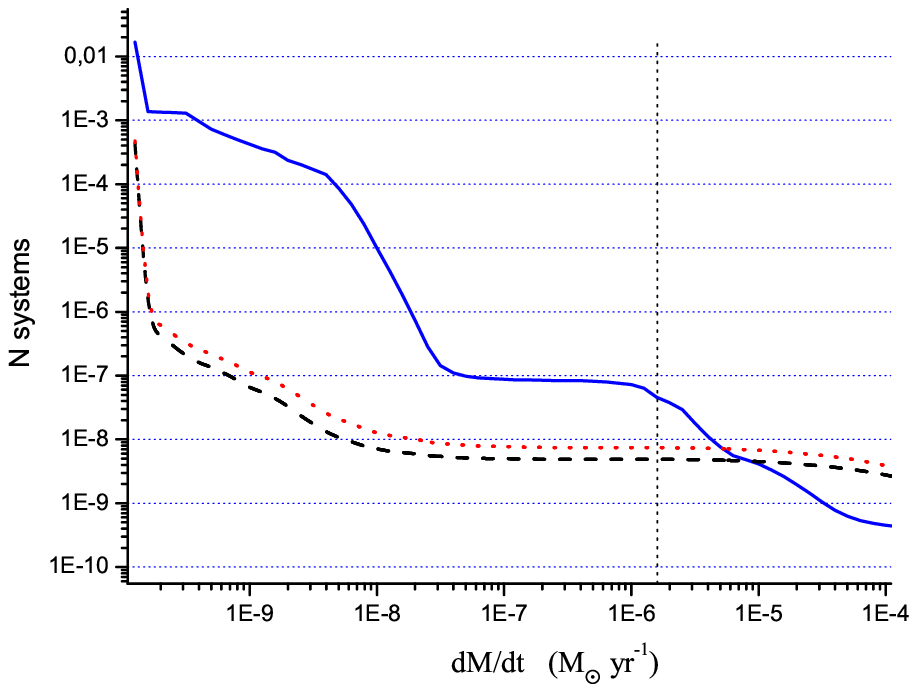}
\includegraphics[width=0.7\textwidth]{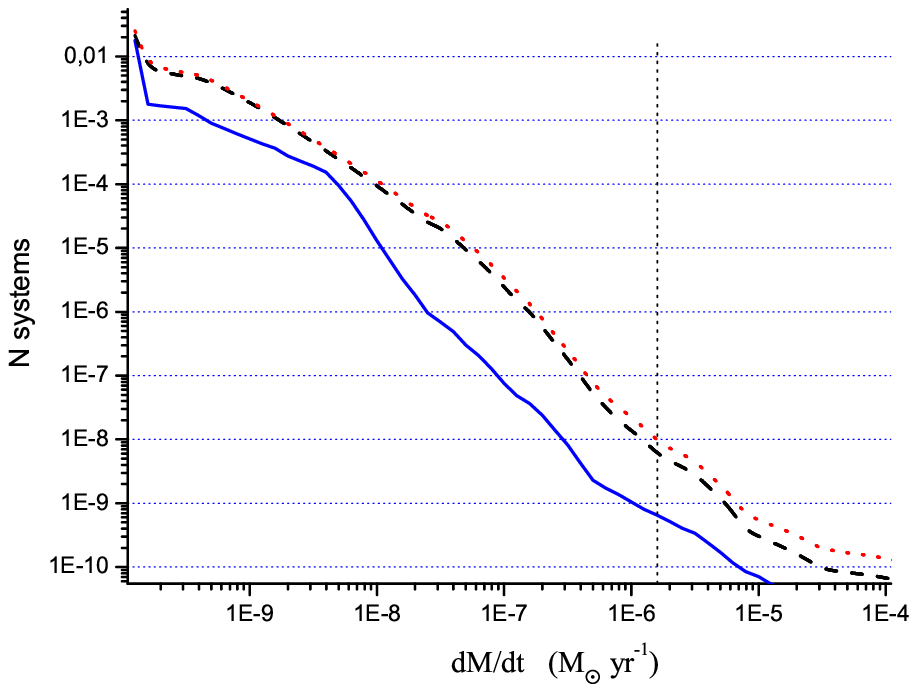}
  \caption{ Integral distribution of the accretion rate for IMBHs formed in the Galactic
  disc from primordial metal-free gas ($\alpha=2.35$). Lines and panels as in Figure \ref{halo35}.}
\label{disk235}
\end{figure}


The tidal capture  of halo IMBHs is a significantly more effective
mechanism of formation of bright accreting binaries.
Still, the number of ULXs formed by means of this channel is found to be very
small, $\sim 10^{-7}$ per galaxy.
Accounting  for low-luminosity sources increases
the number of accreting IMBHs (but not ULXs)
to $\la 0.01$ per galaxy (see Figs. \ref{halo35}, \ref{halo30}).

As in the case of exchange interactions, we also performed
calculations where, instead of the very short Kelvin-Helmholtz timescale of
the red giant envelope, we took the He-burning time in the stellar
core.  Results are not changed significantly.

For some ranges of $\dot M$, in our calculations we found only a few accreting
systems. As we present here integral distributions, this results in
the appearance of ``plateaus''. This is clearly seen in the figures:
the lines for tidal capture often demonstrate a ``step-like'' shape.

For tidal captures, we find three main types of accreting components
to IMBHs.  Two of them are the same as in the case of exchange
interactions (red giants and white dwarfs).  For low accretion rates
$\dot M<3 \times 10^{-9}\, M_{\odot}$~yr~$^{-1}$, companions are
mainly white dwarfs.  For high accretion rates, $ \dot M> 10^{-6}\,
M_{\odot}$~yr~$^{-1}$, most companions are red giants.  For the
intermediate range $3\times 10^{-9}<\dot M<3\times 10^{-8}\,
M_{\odot}$~yr~$^{-1}$ the accreting components can be dwarf main
sequence stars.  In these systems accretion occurs because of orbital
decay due to gravitational wave emission. This mechanism of binary
orbital angular momentum removal is effective for very close binaries
which can be formed after tidal captures.  The mass accretion rate in
this case is determined by the orbital decay timescale.

\subsection{Disc IMBHs}

IMBHs formed in the disc have much lower velocity
dispersion ($\sim  10$ km s$^{-1}$) than halo IMBHs,
so binaries formed in the galaxy plane due to tidal captures by them will have larger
semimajor axes than
those formed from captures by high-velocity halo IMBHs. This results in less effective
ULX formation, see Figs. \ref{disk235}.

\section{Discussion and conclusions}

In this paper we have demonstrated that the probability to form an ULX
by IMBHs wandering in the galactic halo through capturing of a stellar
companion is very low.  There are three reasons for that (see
Eq. \ref{threeprobs}).  First, the very probability of capturing a
stellar companion by IMBHs in the galactic disc is extremely small.
For halo IMBHs it is mainly due to their high velocities during the
galactic disc crossings.  Captures are more probable in the central
regions of galaxies with higher stellar density, but still their
numbers are insufficient to explain the observed ULX statistics in
galaxies.  Another reason is that the bright accretion stage is
usually very short (see also \citep{Blecha&al06}).  The third reason is
that to form an ULX, it is usually necessary to capture a component at
a particular stage of its evolution, when the orbital separation is
small, such as, for example, a very close binary with a white dwarf, or a
system just before coalescence.  For disc IMBHs evolved from
zero-metallicity stars the same three arguments are valid.

Taken together, the three processes listed above, (1) capturing of any
component, (2) short duration of the ULX stage, and (3) small
probability to find a star which is going to be captured on a specific
evolutionary stage, reduce the number of observable ULXs by
$\sim 9 $ orders of magnitude.

We have quantified the relative importance of the three processes and
found that it is the short lifetime of an ULX that mostly reduces the
rate of occurrence of ULXs (by four orders of magnitude alone).  The
necessity to capture a stellar component at the particular evolutionary
stage reduces the ULX occurrence rate by three orders of magnitude.
The dynamical capture probability causes another two orders of
magnitude reduction.

Suppose that in every passage through the galactic plane an
IMBH always captures a star through exchange interactions with binaries.
Not all of the binaries formed this way can later
appear as ULXs.  Let us forget about this for a moment and assume
that the initial binary system always provides the IMBH with a stellar
companion with appropriate properties. Then let us assume that once an
ULX is formed, it is visible for a very long time. Only under such
overly optimistic assumptions the number of IMBHs formed
in  3$\sigma$ halos ($126$ per galaxy) would be  sufficient to produce
the observed $\sim$0.1 ULX per galaxy.

We conclude that the mechanisms of exchange interaction and of tidal
capture cannot explain the observed population of ULXs under the
assumption that they are related to IMBHs wandering in the galactic
halos or formed \textit{in situ} from zero-metallicity stars in
galactic discs.  We have showed that the formation rate of accreting systems
with IMBHs is very low even for smaller luminosities. For
$L>10^{36}$~erg~s$^{-1}$ we obtain only $<0.01$ accreting IMBH per
galaxy.  Hence we conclude that dynamical captures of companions by IMBHs
remnants of zero metallicity stars do not lead to a potentially
interesting number of observable sources.

A much more promising way of forming ULXs from IMBHs can be
achieved in a scenario in which the formation of IMBHs derives
from runaway merging of stars in dense star clusters. In this case
the probability of capturing a companion is very large. The
formation of persistent ULXs is however not granted.
\cite{Blecha&al06} find that IMBH binaries reach a steady ULX
luminosity only for a short period of time ($\simeq 10^5$~years),
making their detection improbable. \cite{Hopman2004},
\cite{bau2005}
 find more optimistic results
for more massive IMBHs, when companion stars are captured through
tidal heating from a mass-segregated population.

Another possibility, which would bypass the need for a stellar
companion, is that the fuel for the IMBHs is provided by a
baryonic remnants that they carry with themselves (see e.g.
\citep{vp2005}). These IMBHs, however, are not to be confused with
"naked" IMBHs ejected from galaxy centers by, e.g., the
gravitational rocket  mechanism. IMBHs accreting from baryonic
remnants are instead associated with inefficient galaxy mergers,
where the IMBHs was hosted by a small satellite falling into the
halo of a large galaxy.  Such light satellites seem to be almost
unaffected by orbital decay, and they are left on peripheral
orbits. The IMBHs hosted in the satellite cores, if accreting,
would be found in the outskirts of a galaxy. \cite{Swartz06},
however, notices that the most (or even all), of ULXs  are found
located within the optical extent of their host galaxies.

\section*{Acknowledgments}
We thank the referee for useful comments.
The work of A.K. and K.P.
was partially supported by the RFBR grants 04-02-16720 and 06-02-16025,
S.P. is the INTAS and Cariplo Foundation Fellow.


\end{document}